\documentclass{ws-p8-50x6-00}

\begin{document}
\date {September 12,1999}
\title{Space Charge In Ionization Detectors\\
And The NA48 Calorimeter\thanks
{Talk presented at CALOR99 --
VIII Int.\  Conf.\ on Calorimetry in HEP,
Lisbon, Portugal, June 13--19,1999}
}

\author{Sandro Palestini\thanks {Representing the NA48 Collaboration: 
Cagliari, Cambridge, CERN, Dubna, Edinburgh, Ferrara, Firenze, Mainz, 
Orsay, Perugia, Pisa, Saclay, Siegen, Torino, Vienna, Warsaw} }

\address{CERN, CH--1211 Geneva 23, Switzerland\\
and INFN - Sezione di Torino, 10125 Torino, Italy\\
E-mail: sandro.palestini@cern.ch}


\maketitle

\abstracts{
The effects of space charge due to slowly drifting ions can be
relevant for detectors operated at high intensity, or for relatively 
low values of the bias voltage. 
Accurate measurements 
have been obtained with the liquid krypton calorimeter of the NA48 experiment, from data collected in 1997.
The build-up of space charge takes place during the first part of the beam extraction burst, and
causes a dependence of the response on the transverse coordinate of the axis of electromagnetic showers, and a small reduction of average amplitude. The effects are well reproduced by a computation, where the only free
parameter is the value of the ion mobility.  
The model can be applied a wide range of operating conditions, 
and generalized to detectors with different geometry and active medium.}
\section{Introduction}
Ionization detectors are usually not affected by ions, because the induced signal is dominated by the fast drifting electrons, 
and the total amount of ion charge accumulated in the detector cell is 
typically much smaller 
than the charge stored on the electrodes. 
However, space charge can produce observable effects in case of
large particle intensities, or 
for relatively low values of the bias voltage.

The liquid krypton calorimeter for the NA48 experiment at CERN was operated 
in 1997 with the bias voltage set at 1.5~kV over a gap of 1~cm, 
half of the value used in the following years.  
In this condition, small but precisely measurable effects of space charge have been observed.

In the following sections, after a brief discussion of the design and
performance of the NA48 calorimeter, a model of space charge effects 
in parallel plate detectors is described, and compared to experimental 
observations. A characteristic, dimensionless parameter is identified,
and predictions over wide ranges of operating conditions, cell geometry, and 
different active media are made. 

\begin{figure}[h]
\begin{center}
\epsfxsize=10.5cm
\epsfbox{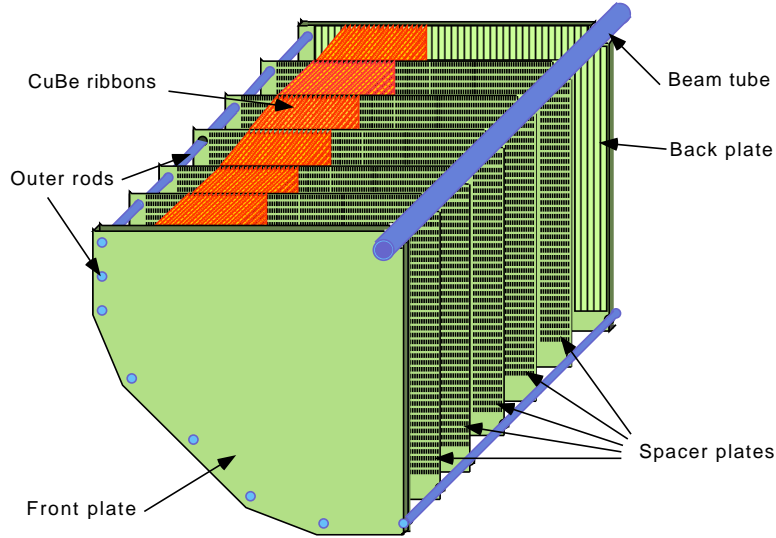} 
\caption{A quadrant of the electrode structure.\label{fig:electrodes}}
\end{center}
\end{figure}
\section{The NA48 liquid krypton calorimeter}
\subsection{Detector design}
\begin{figure}[h]
\begin{center}
\epsfxsize=6cm
\epsfbox{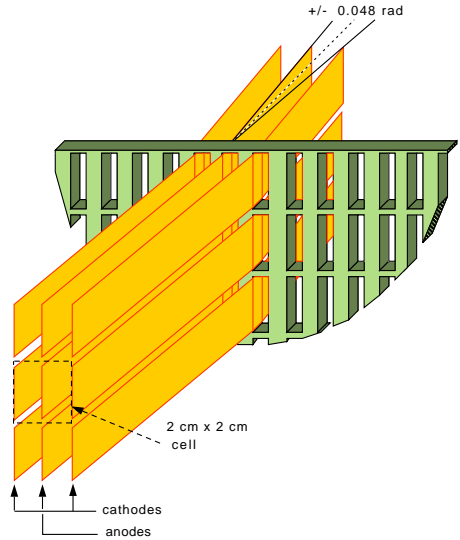}
\end{center}
\caption{Detail of ribbon electrodes near a spacer plate.
\label{fig:dett-electrodes}}
\end{figure}

The NA48 electromagnetic calorimeter is a {\em quasi-}homogeneous 
device based on liquid krypton.\footnote
{Liquid krypton properties: density 2.41~g/cm${^3}$, radiation length 4.7~cm, 
Moli\`ere radius 4.7~cm, boiling point 119.8~K at 1~atm, 
electron drift velocity 
0.27 (0.33) cm/$\mu$s at 1.5 (3)~kV/cm, dielectric constant 
$\epsilon_R=1.7\,$.}
Figure~\ref{fig:electrodes} sketches the electrode structure.
The active surface extends from the beam pipe (8~cm radius) to an octagonal 
outer
boundary 256~cm wide. The read--out cells ($2 \times 2$~cm$^2$) are formed by 
Cu--Be ribbons, 1.8~cm wide and 40~$\mu$m thick. 
The ribbons are stretched between 
front and back plates made of fiber-glass reinforced epoxy, 127~cm apart, and
follow an accordion geometry defined by five thin spacer plates, which guide the 
electrodes through machined slots at $\pm \, 48$~mrad from the vertical plane
(figure~\ref{fig:dett-electrodes}).
The signals induced on each anode are amplified, shaped to 80~ns FWHM wide
signals, and digitized at 40~MHz. For the energy measurement, 
the sum of the signals from a cluster of about 95 cells is used. 

\subsection{Detector performance}
\begin{figure}[h]
\epsfxsize=6cm
\epsfbox{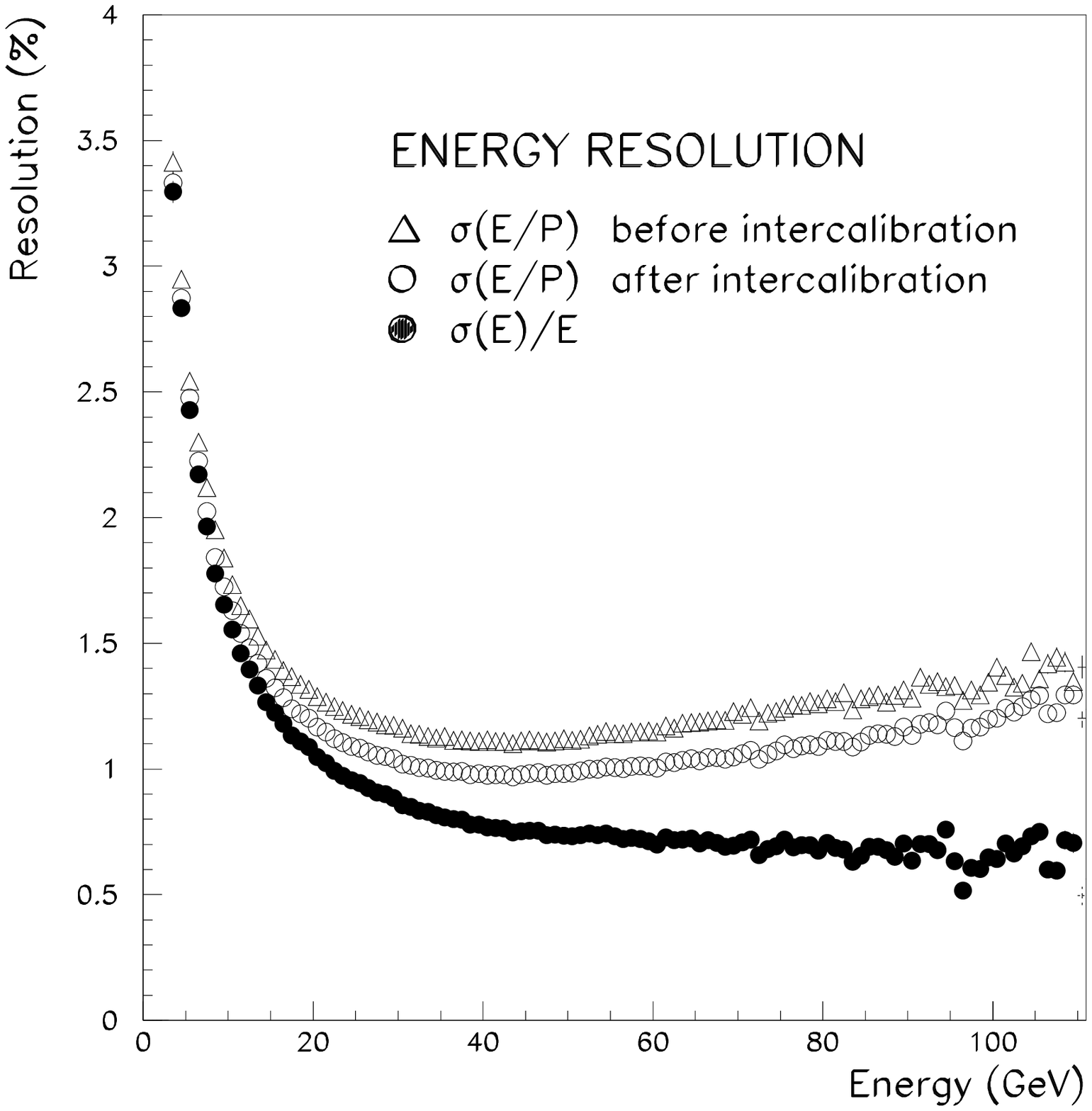}
\epsfxsize=6cm
\epsfbox{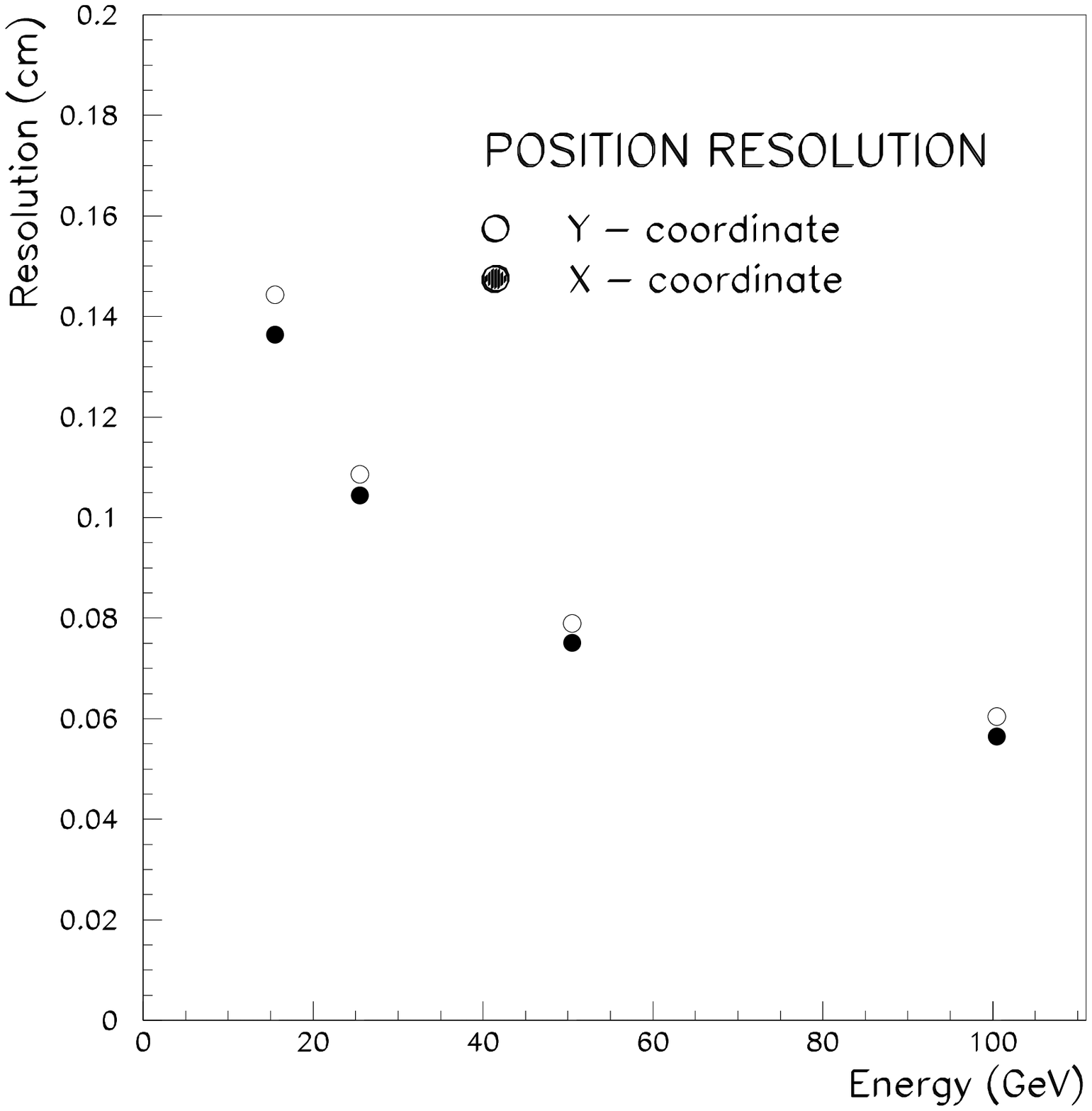}
\caption{Energy and position resolution as a function of the shower energy.
The lowest set of data points in the figure on the left is obtained after
unfolding the spectrometer contribution and provides the calorimeter 
resolution.
\label{fig:performance}}
\end{figure}

The performance of the detector has been obtained using electrons
from semileptonic kaon decays, for which the momentum $P$ is measured 
in a magnetic spectrometer, and with special runs 
taken with electron beams. 
As shown in figure~\ref{fig:performance},
the energy resolution is
$\sigma (E) /E \simeq 0.125/E \oplus 0.032/\sqrt{E} \oplus 0.005$  
($E$ in GeV),
where the different terms are added in quadrature. 
The spatial resolution (figure~\ref{fig:performance})
is better than 1.3~mm and the time resolution 
is better than 300~ps for showers above 20~GeV.

A discussion of additional detector characteristics particularly 
relevant for the NA48 physics program can be found in reference 1. 

\section{Space charge model}
\subsection{General considerations and assumptions}
In order to develop a formalism to describe the effects of space charge, 
the following observations and assumptions are made: 
\begin{itemize}
\item the drift velocity of the positive ions is much lower 
than the electron velocity ($v_e$), so that only ions effectively 
contribute to space charge ($\rho$);
\item the ion drift velocity is determined by the mobility coefficient $\mu$, 
which is assumed constant;
\item the ionization detector is approximated as an ideal parallel plate 
capacitor (with the electric field $E$ along the $x$ direction);
\item for the moment, we assume that the injected charge density rate $J$
is uniform inside the cell.
\end{itemize}

Under the assumptions above, $\rho$ and $E$ must satisfy the 
charge-con\-ser\-va\-tion equation:
\begin{equation} \label{eq:timedep}
\frac{\partial \rho}{\partial t} \,+\, \mu\,E\,
\frac{\partial \rho}{\partial x}\,+\,
\frac{\mu}{\epsilon}\,\rho^2\:=\:J
\end{equation}
where the time evolution of the space charge 
is related to the the ion drift from anode to cathode (second term), 
to a term quadratic in $\rho$ (related to  $\partial E/\partial x$, 
$\epsilon$ is the dielectric constant of the medium), 
and to the charge density injection rate.

\subsection{The stationary case}
The stationary solution can be written as:
\begin{equation}
\rho = \frac {J\, x}{\mu \, E}
\end{equation}
\begin{equation}\label{eq:E_scaling}
E \: = \: \frac{V}{X} \sqrt{C^2+\alpha^2\frac{x^2}{X^2}}
\end{equation}
where $V$ is the value of the bias voltage, $X$ is the width of the cell
($x=0$ at the anode and $x=X$ at the cathode), $C$ is a 
numerical integration constant, and the dimensionless  parameter $\alpha$ is
defined as:
\begin{equation} \label{eq:alpha}
\alpha  =  \frac{X^2}{V}\sqrt{\frac{J}{\epsilon\,\mu}}
\end{equation}

\begin{figure}[h]
\begin{center}
\epsfxsize=7.95cm
\epsfbox{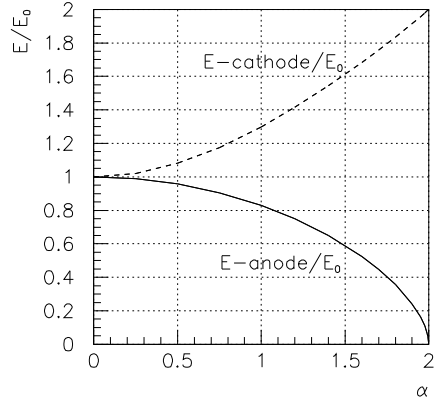}
\end{center}
\caption{Normalized electric field at anode and cathode vs.\ $\alpha$.
\label{fig:C}}
\end{figure}
The constant $C$ 
corresponds to the electric field at the anode, divided  by $E_0=V/X$. 
Figure~\ref{fig:C} shows the value of $C$ as function of $\alpha$,
together with the corresponding values of $E/E_0$  at the cathode. 
From this and from equation~\ref{eq:E_scaling}, 
we see that once dimensionless variables $x/X$, $E/E_0$ are used,
the problem is completely defined by the value of the parameter $\alpha$.
The distortion to the electric field due
space charge can be controlled, with increasing degrees of 
sensitivity, by changing the values of $J$, $V$, and $X$.\footnote
{Modifications are necessary in order to deal with very large space charge
density
occurring for $\alpha > 2$. A discussion of this, and of some other aspects of
space charge effects, can be found in reference~2.} 

\begin{figure}[th]
\begin{center}
\epsfxsize=12.4cm
\epsfbox{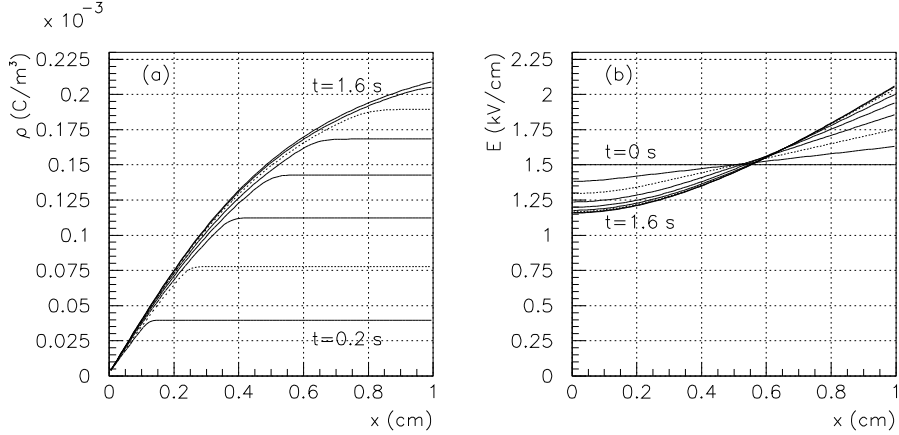}
\end{center}
\caption{(a) Space charge density and (b) electric field vs.\ $x$, for 
different time values during the charge injection interval, 
for $\alpha=1.15\,\,$. \label{fig:timedep}}
\end{figure}
\subsection{Time dependence}
Of particular interest is the case when the charge injection rate $J$ vanishes 
for $t < 0$, and is constant for $t > 0$. A numerical integration of 
equation~\ref{eq:timedep} is shown in figure~\ref{fig:timedep}, which provides
the behavior of $\rho $  and $E$ as functions of $x$, 
at various values of $t$. 
The following values have been used for the various parameters:
$\mu$=0.45$\times 10^{-7}\,$m$^2$V$^{-1}$s$^{-1}$, 
$\epsilon$=1.5$\times 10^{-11}$~F m$^{-1}$ 
(corresponding to liquid krypton), $V$=1500~V, $X$=1.0~cm,
and $J = 2.0\times 10^{-4}$~C~m$^{-3}$s$^{-1}$ for $t > 0$.
The corresponding value for $\alpha$ is $1.15\,$. 
The different lines for $\rho$ and $E$ 
correspond to values of $t$ in steps of 0.2~s. 
The computation shows that 
a stationary solution is reached for time larger than about 1.5~s, 
which is equal to the maximum drift time for ions (at low intensity): 
$X^2/(\mu V)$.  

\begin{figure}[th]
\begin{center}
\epsfxsize=12cm
\epsfbox{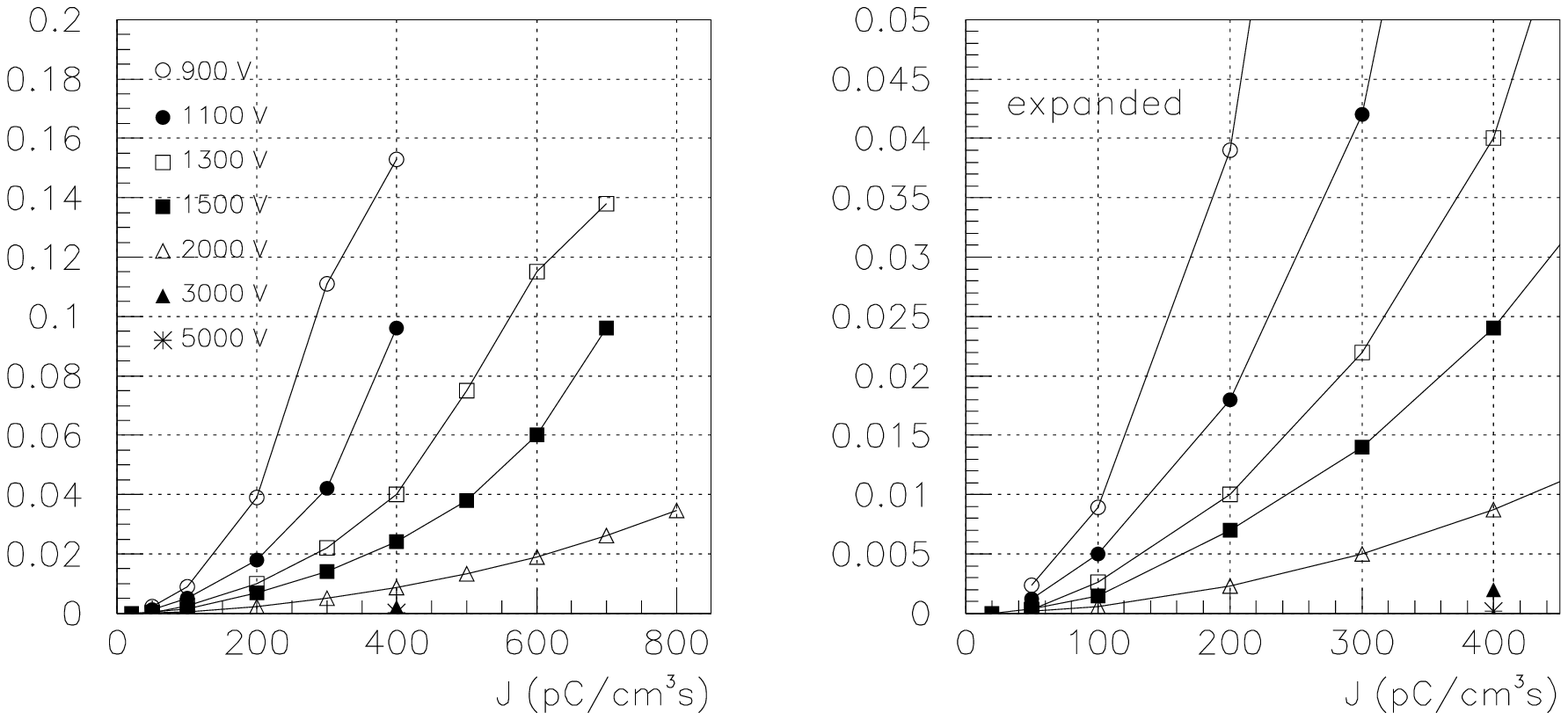}
\epsfxsize=12cm
\epsfbox{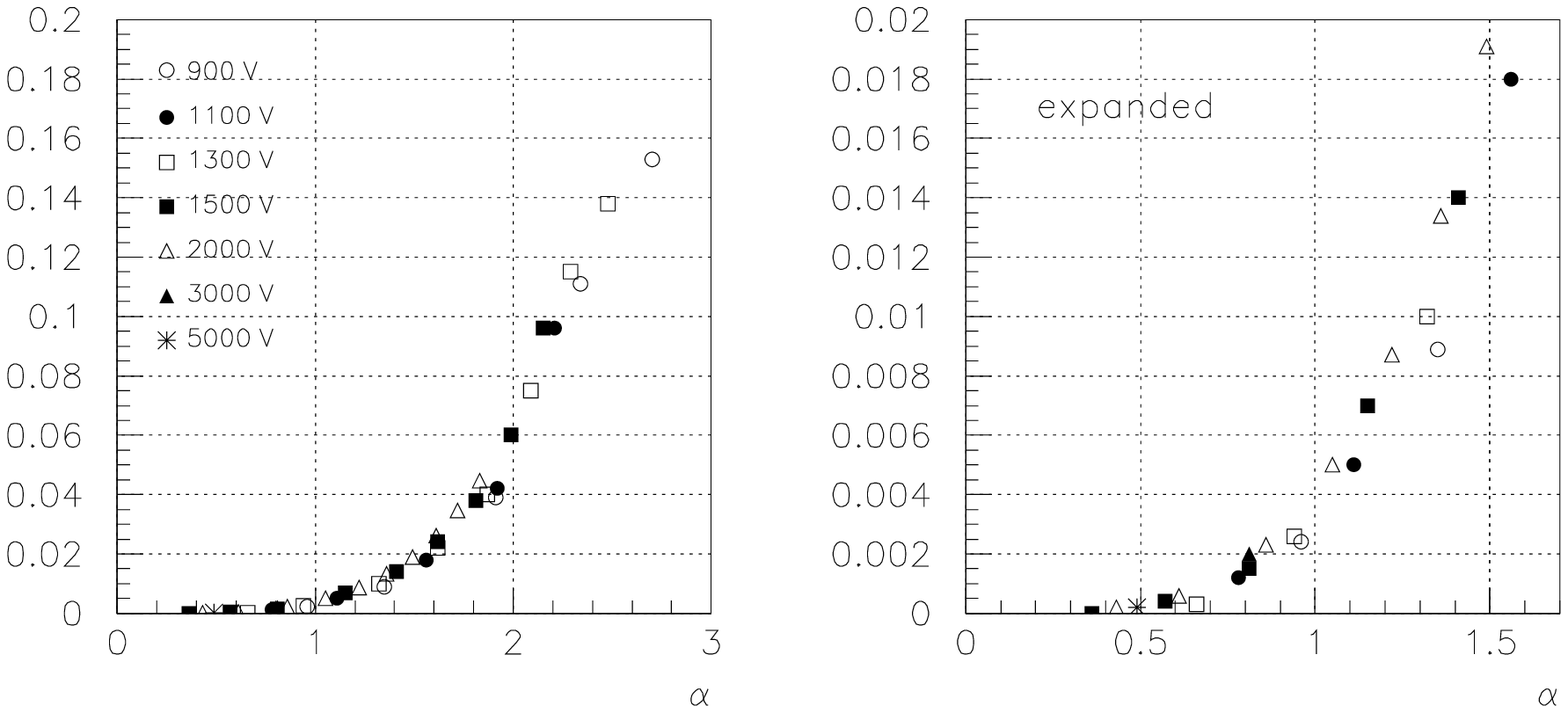}
\end{center}
\caption{Top: Fractional reduction of average response 
vs.\ charge injection rate for liquid krypton,
with $X=1$ and different voltage values. Bottom: as above, but as a function
of the parameter $\alpha$. \label{fig:ave-red}}
\end{figure}
\section{Ionization detectors}
In ionization detectors, the charge deposited at the position $x$
induces on the electrode a current proportional to the
electron drift velocity $v_e(x)$, 
which depends on the local value of the electric field $E(x)$ 
through the relation between $v_e$ and $E$. 
Space charge due to positive ions decreases the response near 
the anode, and increases it near the cathode, following the trend of $E(x)$.  At low intensity ($\alpha \leq 0.5$), the variation in 
$v_e$ may result to good approximation proportional 
to the variation in $E$ ({\em i.e.}: $\delta v_e/v_e \propto \delta E/E$), 
and the average of the response across the cell width
(the {\em average response\/}) is unchanged: 
$<\!v_e\!>_x = v_e(<\!E\!>_x) = v_e(E_0)$.
For larger values of $\alpha$ the convexity of $v_e(E)$ is such 
that the increase in $v_e$ near the cathode does not match the 
decrease at the anode, and the average response of an ionization detector 
is reduced by the presence of space charge. 

\begin{figure}[t]
\begin{center}
\epsfxsize=7.2cm
\epsfbox{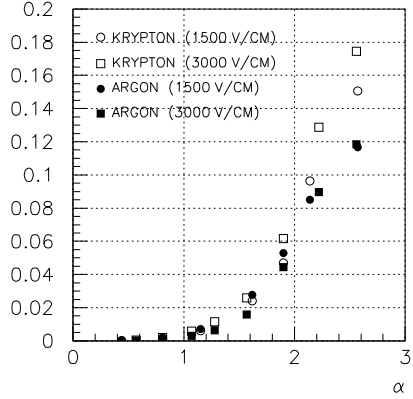}
\end{center}
\caption{Average reduction of response vs.\ $\alpha$ on liquid krypton and
liquid argon. \label{fig:kry-arg}}
\end{figure}
\begin{figure}[h]
\begin{center}
\epsfxsize=12cm
\epsfbox{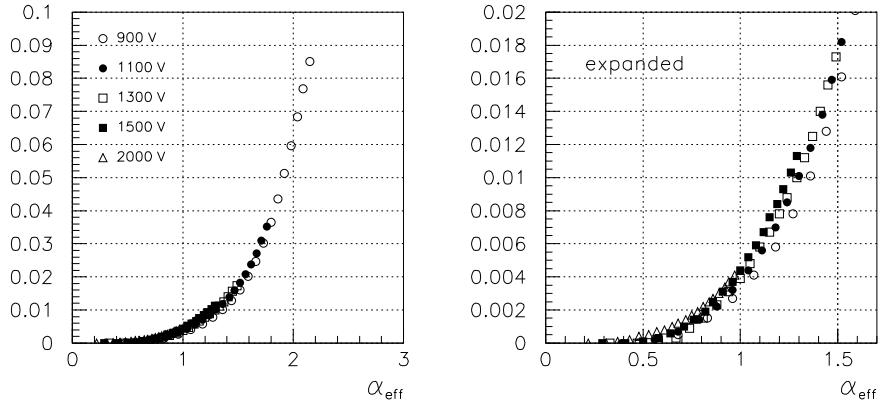}
\end{center}
\caption{Decrease of average response for showers in krypton, as a function 
of $\alpha$. \label{fig:showers}}
\end{figure}

Using accurate parametrizations of 
$v_e(E)$,\cite{v-drift,v-drift-argon}
it is found that to good approximation the reduction in 
average response depends only on the value of the parameter $\alpha$, 
as shown in figure~\ref{fig:ave-red} 
for krypton, and in figure~\ref{fig:kry-arg} for krypton and 
argon together.\footnote
{As for figure~\ref{fig:timedep}, the computations shown in 
figures~\ref{fig:ave-red}--\ref{fig:showers}
have been made with the cell width $X=1$~cm.}
  We shall refer to this as {\em approximate
scaling\/} of the reduction of the average response. 

A similar situation occurs for calorimeters. 
The effective charge density injection
rate $J_{\mathrm{eff}}$ is proportional to the energy flux into the detector, 
divided by the average ionization energy and by the detector
effective depth. The latter depends on the shower longitudinal
profile, and for fully absorbed 
electromagnetic showers, is equal to 13 radiation lengths. 
Figure~\ref{fig:showers} shows that approximate scaling is
still valid, with the reduction of response depending on
$\alpha_{\mathrm{eff}}=\alpha(J_{\mathrm{eff}})$ as in equation \ref{eq:alpha}. The formalism describing the average response is quite general, 
and can be applied to detectors 
with different electrode structure, and sampling calorimeters.

\section{Comparison with measurements}
\begin{figure}[h]
\epsfxsize=6cm
\epsfbox{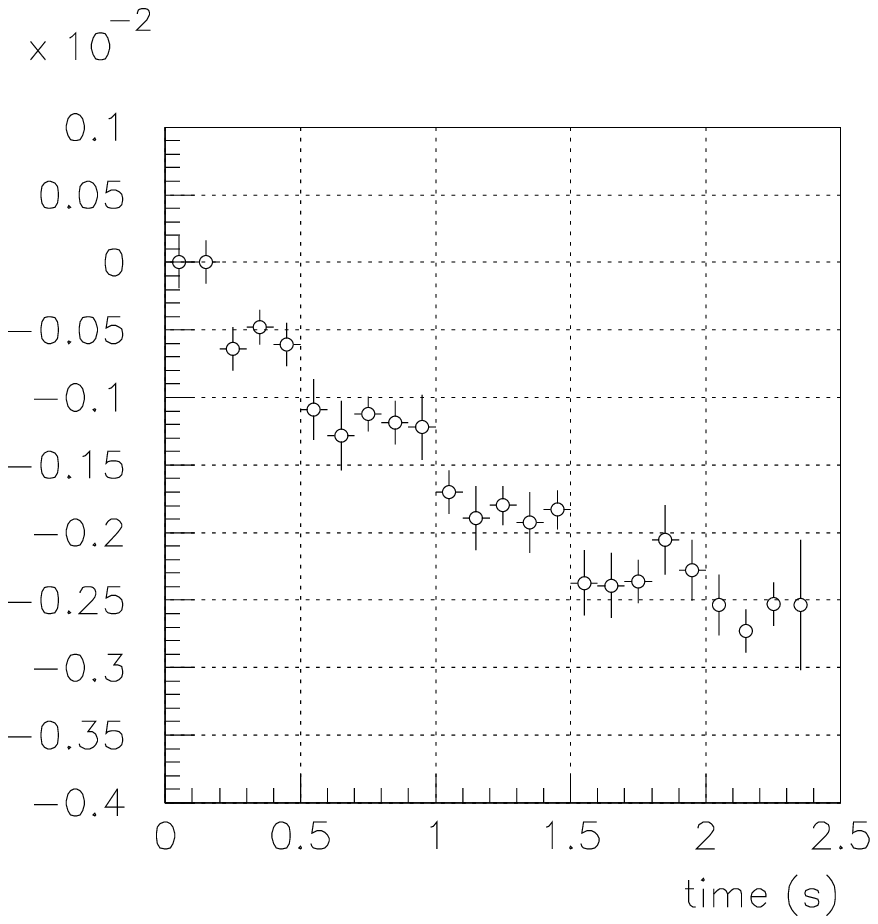}
\epsfxsize=6cm
\epsfbox{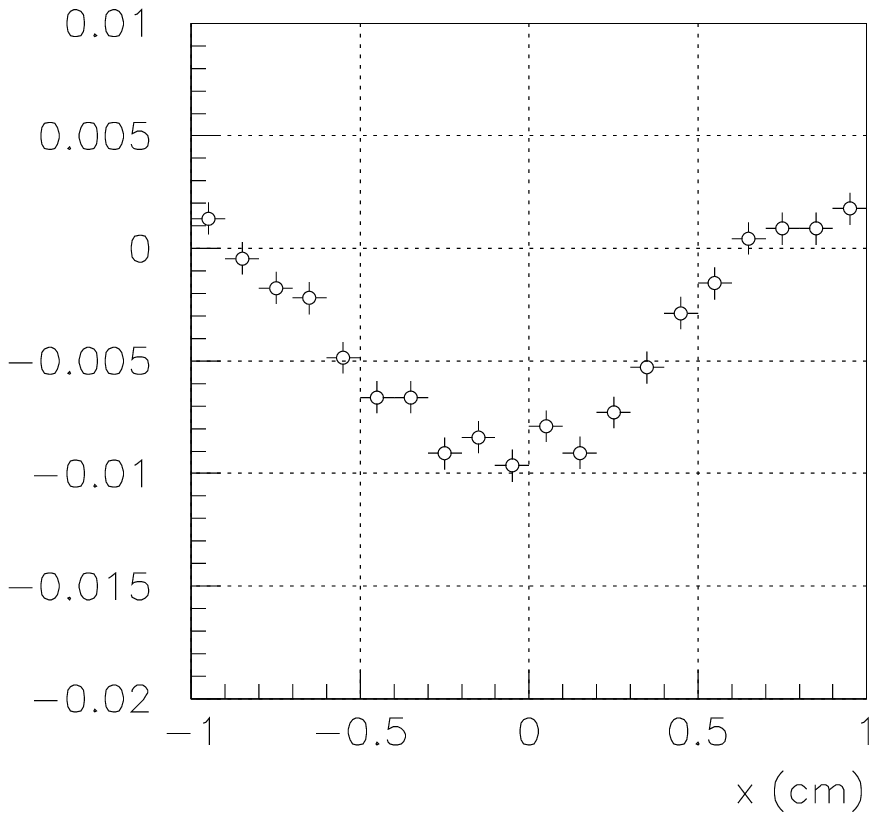}
\caption{Left: dependence of the average response on the time from the 
beginning of the extraction period ($r<30$~cm). 
Right: dependence of the local response 
on the shower coordinate within the central read--out cell
($20<r<25$~cm). The value $x=0$ 
corresponds to the anode, and $x=\pm \, 1$ to the cathodes.
\label{fig:meast-x}}
\end{figure}

Figure~\ref{fig:meast-x}(left) shows the reduction in average response,
as a function of the time within the beam extraction period,  observed with the NA48 calorimeter, operated at 1.5~kV, 
for showers with radial position $r < 30$~cm.  
The measurement is performed with electrons from semileptonic
kaon decays, comparing the calorimeter energy measurement 
to the momentum measured in the magnetic spectrometer.
The observed trend is compatible with the value 
$\mu = (0.41 \pm 0.02)$~cm$^2$kV$^{-1}$s$^{-1}$ for the ion mobility.  
Comparison of response between early showers 
($t < 0.1$~s) and late ones ($t > 1.6$~s) provides the {\em local\/} 
variation of response related to $E(x)$, shown in 
figure~\ref{fig:meast-x}(right) for showers with $20 < r < 25$~cm. 

A comparison between measurements and model predictions 
can be made based on the mobility value quoted above, and
known values for the charge injection rate. 
Across the NA48 detectors, the particle intensity varied in 1997 from 
1500~GeV cm$^{-2}$s$^{-1}$ at $r = 15$~cm ($\pm \, 5 \%$), 
to a value 100 times smaller at $r = 120$~cm. 
The corresponding maximum value 
for the effective charge injection rate and scaling parameter 
are $J = 180$~pC cm$^{-3}$s$^{-1}$ and $\alpha_{\mathrm {eff}} = 1.1\,$.

\begin{figure}[h]
\begin{center}
\epsfxsize=7.8cm
\epsfbox{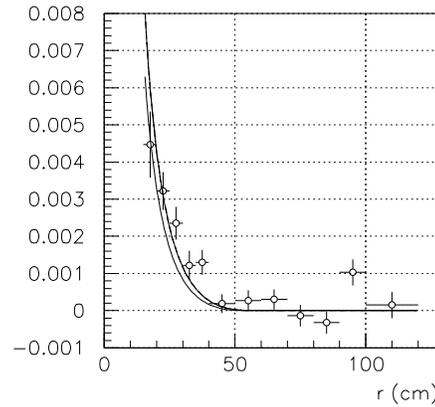}
\caption{Measured reduction of average response (data points) compared to the 
model prediction.
\label{fig:comparison}}
\end{center}
\end{figure}
Figure~\ref{fig:comparison} shows the comparison between data 
and model predictions. 
The two lines define the range predicted by the model 
given the uncertainties in the intensity 
normalization and in the ion mobility.  It should be stressed that 
the model is not a fit to the data, and its shape and normalization 
have been obtained from general principles and independent measurements. The good agreement is therefore relevant and 
establishes confidence in the various assumptions used in this analysis.

\begin{figure}[h]
\begin{center}
\epsfxsize=11.52cm
\epsfbox{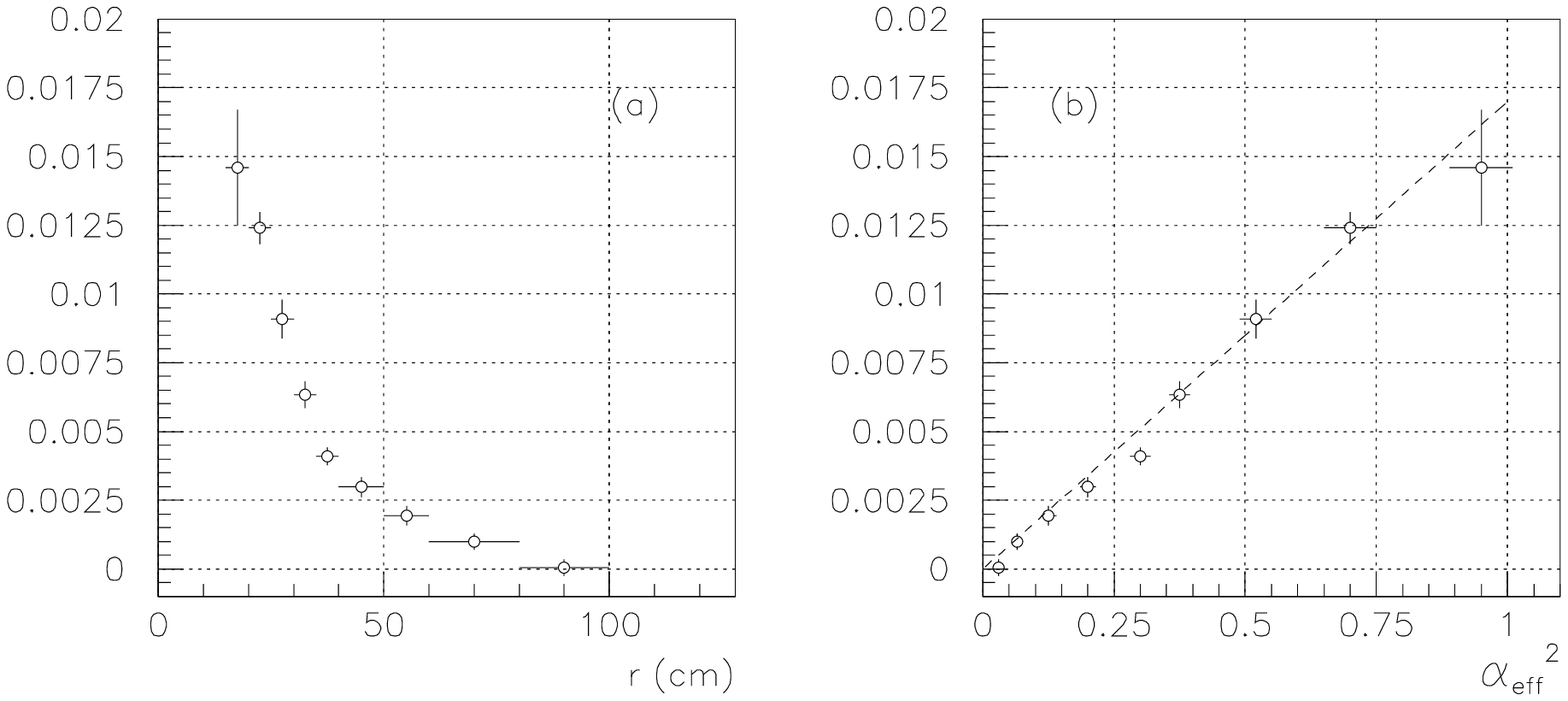}
\caption{Dependence of the peak-to-peak local modulation of the response 
on the radial position (a), and on the parameter $\alpha_{\mathrm{eff}}{}^2$ 
(b).\label{fig:peak-to-peak}}
\end{center}
\end{figure}
Figure~\ref{fig:peak-to-peak} shows the peak-to-peak amplitude of the 
$x$-dependent (local) non-uniformity of response, as function of 
the $r$ and of $\alpha_{\mathrm {eff}}{}^2$.  
The latter is characterized by  a predictable  
linear relation.\footnote
{The proportionality coefficient depends on the ratio between 
cell width and Moli\`ere radius, and its value will change for 
detectors with different read--out cell or medium.} 

\section{Conclusions}
Effects caused by space charge due to positive ions have been
observed with the liquid krypton calorimeter of the experiment NA48. A model has been developed, characterized by  a dimensionless parameter, and suited to describe a wide range of operating conditions and different detector designs.  Good agreement is found between experimental data and model predictions.

\end{document}